\RequirePackage{lineno}
\documentclass[aps,prc,twocolumn,superscriptaddress,linenumbers,nofootinbib]{revtex4}
\usepackage{graphicx}
\usepackage{gensymb}
\usepackage{subfigure}
\usepackage{pdfpages}
\usepackage[normalem]{ulem}
\makeatletter
\AtBeginDocument{\let\LS@rot\@undefined}
\makeatother

\usepackage{amsmath}
\newcommand{\GeVsq}[0]{$({\rm GeV}\!/\!c)^2$}
\usepackage{hyperref}  
\hypersetup{breaklinks = true, colorlinks = true, allcolors = blue}
\usepackage{multirow}

\begin{document}



\title{Polarization transfer to
bound protons \\ measured by quasi-elastic electron scattering on $^{12}$C}
\newcommand{\affJSI}[0]{\affiliation{Jo\v{z}ef Stefan Institute, 1000 Ljubljana, Slovenia.}}
\newcommand{\affTAU}[0]{\affiliation{School of Physics and Astronomy, Tel Aviv University, Tel Aviv 69978, Israel.}}
\newcommand{\affJGU}[0]{\affiliation{Institut f\"ur Kernphysik, Johannes Gutenberg-Universit\"at, 55099 Mainz, Germany.}}
%
\author{Tilen Brecelj$^{\ddagger}$}
\renewcommand*{\thefootnote}{\fnsymbol{footnote}}
\footnotetext[3]{These authors contributed equally to this work}
\renewcommand*{\thefootnote}{\arabic{footnote}}

\affJSI
\author{Sebouh J. Paul$^{\ddagger}$}
\email[Corresponding author: ]{paulsebouh@mail.tau.ac.il}

\thanks{\\$^\ddagger$These two authors contributed equally to this work.}
\affTAU
\author{Tim Kolar}\affJSI
\author{Patrick Achenbach}
\affJGU
\author{Adi Ashkenazi}
\affTAU
\author{Ralph B\"ohm}
\affJGU
\author{Erez O. Cohen}
\affTAU
\author{Michael O. Distler}
\author{Anselm Esser}
\affJGU
\author{Ronald Gilman}
\affiliation{Rutgers, The State University of New Jersey, Piscataway, NJ 08855,
USA.}
\author{Carlotta Giusti}
\affiliation{Dipartimento di Fisica, Universit\`a degli Studi di Pavia and INFN, Sezione di Pavia, via A.~Bassi 6, I-27100 Pavia, Italy.}
\author{David Izraeli}
\affTAU
\author{Igor Korover}
\affiliation{Department of Physics, NRCN, P.O. Box 9001, Beer-Sheva 84190, Israel.}
\affTAU
\author{Jechiel Lichtenstadt}
\affTAU
\author{Israel Mardor}
\affTAU
\affiliation{Soreq NRC, Yavne 81800, Israel.}
\author{Harald Merkel}
\affJGU
\author{Miha Mihovilovi\v{c}}
\affiliation{Faculty of Mathematics and Physics, University of Ljubljana, 1000
Ljubljana, Slovenia.}
\affJSI
\affJGU
\author{Ulrich M\"uller}
\affJGU
\author{Mor Olivenboim}
\author{Eli Piasetzky}
\affTAU
\author{Guy Ron}
\affiliation{Racah Institute of Physics, Hebrew University of Jerusalem, Jerusalem
91904, Israel.}
\author{Bj\"orn S. Schlimme}
\author{Matthias Schoth}
\author{Florian Schulz}
\author{Concettina Sfienti}
\affJGU
\author{Simon \v{S}irca}
\affiliation{Faculty of Mathematics and Physics, University of Ljubljana, 1000
Ljubljana, Slovenia.}
\affJSI
\author{Samo \v{S}tajner}
\affJSI
\author{Steffen Strauch}
\affiliation{University of South Carolina, Columbia, South Carolina 29208, USA.}
\author{Michaela Thiel}
\author{Adrian Weber}
\affJGU
\author{Israel Yaron}
\affTAU

\collaboration{A1 Collaboration}

\date{\today}

\begin{abstract}
We report the measurements of the transverse ($P'_x$) and longitudinal ($P'_z$) components of the polarization transfer to a bound proton in carbon via the quasi-free $^{12}{\rm C}(\vec e,e'\vec p)$ reaction, over a wide range of missing momenta.  We determine these polarization-transfers separately for protons knocked out from the $s$- and $p$-shells.
The electron-beam polarization was measured to determine the individual components with systematic uncertainties which allow a detailed comparison with theoretical calculations.
\end{abstract}


\maketitle

\section{Introduction}
Polarization transfer from a polarized electron to a proton in elastic scattering has
become a recognized method to measure the proton's elastic electromagnetic form factors, $G_E$ and $G_M$ \cite{Jones:1999rz,Gayou:2001qd,Punjabi:2005wq,Milbrath:1997de,Barkhuff:1999xc,Pospischil:2001pp,PhysRevC.64.038202,MACLACHLAN2006261,PhysRevC.74.035201}.   Assuming the one-photon exchange approximation, the ratio of the transverse ($P'_x$) to longitudinal ($P'_z$) polarization-transfer components is proportional to the ratio of these form factors, $G_E/G_M$ \cite{Akh74}.
This provides a direct measurement of the form factor (FF) ratio, even under conditions where one of the FFs is much larger than the other, and eliminates many systematic uncertainties \cite{Perdrisat}.  

Measuring the ratio of the components of the polarization transfer to a \textit{bound} proton in \textit{quasi-free}
kinematics on nuclei, which is sensitive to the electromagnetic FF ratio,
has been suggested as a method to study differences between free and bound protons \cite{Milbrath:1997de,Barkhuff:1999xc}.  
As such it can be used as a tool to identify
medium modifications in the bound proton's internal structure, reflected in the FFs and thereby in the
polarization transfer.  Deviations between measured polarization ratios in quasi-free $A(\vec e,e'\vec p)$ and elastic $\vec ep$ scattering can be interpreted only by comparing the measurements with realistic calculations of nuclear effects such as final-state interactions (FSI).

Polarization-transfer experiments have been carried out on $^2$H and $^{12}$C target nuclei at the
Mainz Microtron (MAMI) \cite{deep2012PLB,deepCompPLB,posPmissPLB,ceepLet}, as well as on $^2$H, $^4$He and $^{16}$O at Jefferson Lab (JLab) \cite{Strauch,jlabDeep,Paolone}, in search of medium modification in the proton internal structure. These experiments were performed to study deeply bound nucleons, characterized by high
missing momentum which is equivalent (neglecting FSI) to protons with high initial momentum.  It was shown for the $^2$H measurements that the deviations in $P'_x/P'_z$ from that of elastic $\vec ep$ scattering can be explained by
nuclear effects without the necessity of introducing modified FFs \cite{deep2012PLB,deepCompPLB,posPmissPLB}.  
Furthermore, when comparing the quasi-elastic polarization transfer to that of elastic $\vec ep$ scattering, the double ratio $(P'_x/P'_z)^A/(P'_x/P'_z)^{^1\!\rm H}$ exhibits a very similar behavior for  $^2$H \cite{deep2012PLB,posPmissPLB,jlabDeep}, $^4$He  \cite{Paolone}, and $^{12}$C \cite{ceepLet}, suggesting a universality.

While the ratio of the components is better determined experimentally than the individual components (smaller systematic uncertainties), it is insensitive to possible common effects to $P'_x$ and $P'_z$ which cancel in the ratio.
To test calculations in better detail, and further corroborate their reliability, measurements of individual components of the polarization transfer are required.   

The polarization-transfer components have previously been measured for $^2$H at MAMI \cite{deepCompPLB,movingProtonPLB,posPmissPLB}, as well as $^4$He at JLab \cite{Paolone}.   
For $^2$H, statistically significant but relatively small deviations were observed between the measured components and the predicted values, but not in their ratio \cite{deepCompPLB,posPmissPLB}.

The $^{12}$C nucleus is a particularly appealing target for such studies as one can selectively probe protons from specific nuclear shells, $s$ and $p$.  The average local densities in these shells differ by about a factor of two, which was predicted to impact the polarization transfer to $s$- and $p$-shell protons differently \cite{PhysRevC.87.028202}.  
If modifications to the bound-proton structure by the nuclear medium exist, and are reflected in the proton FFs, then they may depend either on the off-shellness of the bound proton (virtuality), or on the local nuclear density, or both.  Therefore, it is important to measure the components of the polarization-transfer in both shells over a large missing-momentum range and to have reliable calculations for them in order to identify modifications which can be related to the FFs.  

In this work, we present measurements of the $^{12}{\rm C}(\vec e,e'\vec p)$ reaction at MAMI over a wide range of missing momentum of the struck proton.   The beam-polarization determination had a sufficient accuracy for extracting the individual polarization-transfer components $P'_x$ and $P'_z$ and for allowing a meaningful comparison with theoretical calculations.  The results of the measured ratio $P'_x/P'_z$ have been reported in \cite{ceepLet}.  
The new analysis presented here includes improvements in the corrections for the energy loss of the particles when exiting the target, as well as the polarization-extraction procedure.

Section \ref{sec:hardware} describes the experimental setup, including the beam-line, target and spectrometers.  Details of the measured reaction and its kinematics are given in Section \ref{sec:kin}.   
The data analysis and determination of the polarization components are described in Section  \ref{sec:measurement}.  
Finally, in Section \ref{sec:interp}, we  interpret the data,  compare them to a set of calculations, and explore the sensitivity to the proton FFs.  

\section{Experimental setup\label{sec:hardware}}

The experiment was performed at MAMI using the beam-line and spectrometers of the A1 Collaboration \cite{a1aparatus}. We used a 600 MeV continuous-wave polarized electron beam with a current of about 10 $\mu$A and an $\sim$80\% polarization.  The beam helicity was flipped at a rate of 1 Hz in order to reduce time-dependent systematics in the polarization-transfer measurements.  

\begin{figure}
\includegraphics[width=\columnwidth]{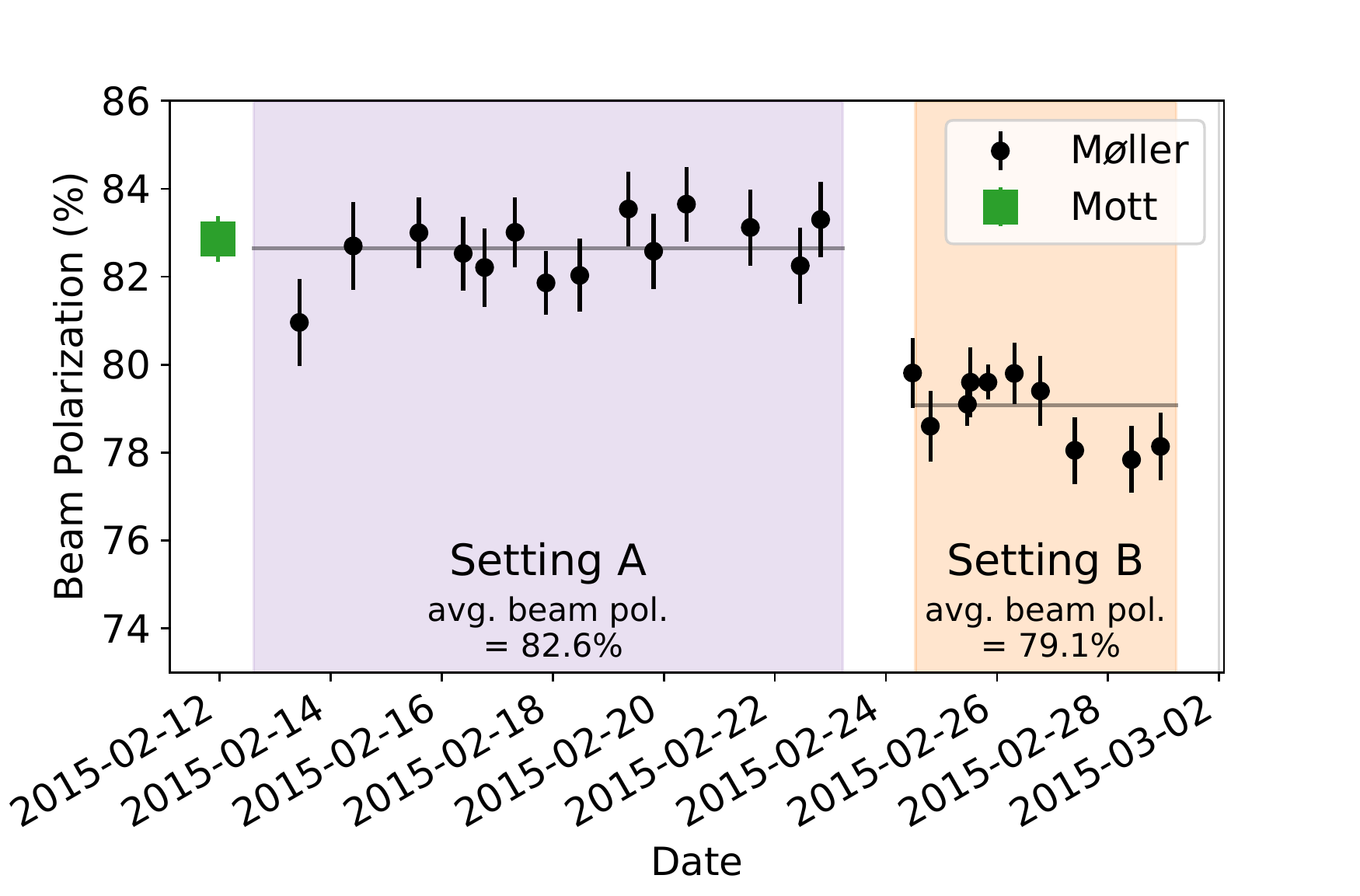}
\caption{
The measured beam polarizations using the M\o ller (black circles) and Mott (square, green online) polarimeters.  The average beam polarizations for each dataset are shown as horizontal lines.  The error bars shown are statistical errors, and do not include systematic errors, estimated to be around 2\% for both the M\o ller and Mott polarimeters.  The shaded regions represent the data-taking periods for the two kinematic settings of the experiment.  
}
\label{fig:beam_pol}
\end{figure}

The beam polarization was measured with a M\o ller polarimeter \cite{Wagner,Bartsch} about twice daily, and verified by a Mott polarimetry measurement \cite{Tioukine} taken at the start of the measurement period.  These two methods of beam-polarization measurement produced mutually consistent results, as shown in Fig.~\ref{fig:beam_pol}, and each had a systematic uncertainty of 2\%.

In between taking the two datasets at different kinematic settings, we refreshed the GaAs crystal in the polarized beam source.  This resulted in a drop in the beam polarization from 82.6\% to 79.1\%.  These values reflect the average beam-polarization measurements before and after the crystal was refreshed.  During the data-taking, the beam polarization remained constant within error.

\begin{figure}[h]
\includegraphics[width=.5\columnwidth]{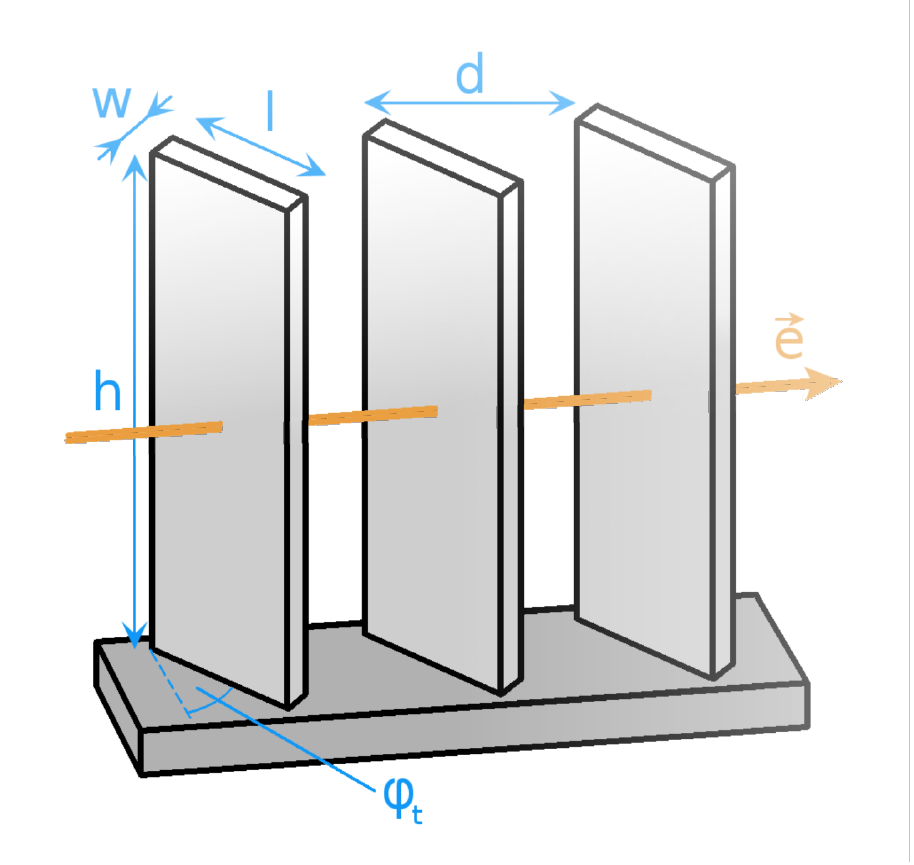}
\caption{
Schematic view of the carbon target, consisting of three graphite foils.  The thick arrow (orange online) indicates the direction of the electron beam.  The dimensions shown are $h\times l \times w=20$ mm $\times$ 4 mm $\times$ 0.8 mm,  $d=15$ mm, and $\phi_t=40\degree$.
}
\label{fig:target_dim}
\end{figure}

The target consisted of three carbon (graphite) foils of 0.8 mm thickness each, separated by about 15 mm and tilted at an angle of 40$\degree$ with respect to the beam.  Their transverse dimensions were 4 mm $\times$ 20 mm, as shown in Fig.~\ref{fig:target_dim}.  This design reduced the distance that the protons would travel through the target foils, both by using multiple foils (rather than a single thick foil) and by rotating each foil such that the detected protons would exit nearly normal to the foil.  This reduced their energy loss in the target.  It also improved the resolution for the reaction-vertex determination, consequently reducing the systematic uncertainty in the measured polarization-transfer components at the reaction point. 

  Two high-resolution, small-solid-angle spectrometers with momentum acceptances of 20-25\% were used to detect the scattered electrons in coincidence with the knocked-out protons.  Each of these spectrometers consists of a magnet system with a quadrupole-sextupole-dipole-dipole configuration, followed by  vertical drift chambers (VDC) for tracking, and a scintillator system for triggering and the timing coincidence between the two spectrometers.  The electron spectrometer also includes a \v Cerenkov detector for identifying electrons and distinguishing them from background particles such as $\pi^-$ and cosmic rays.

\begin{figure}[h]
\includegraphics[width=\columnwidth]{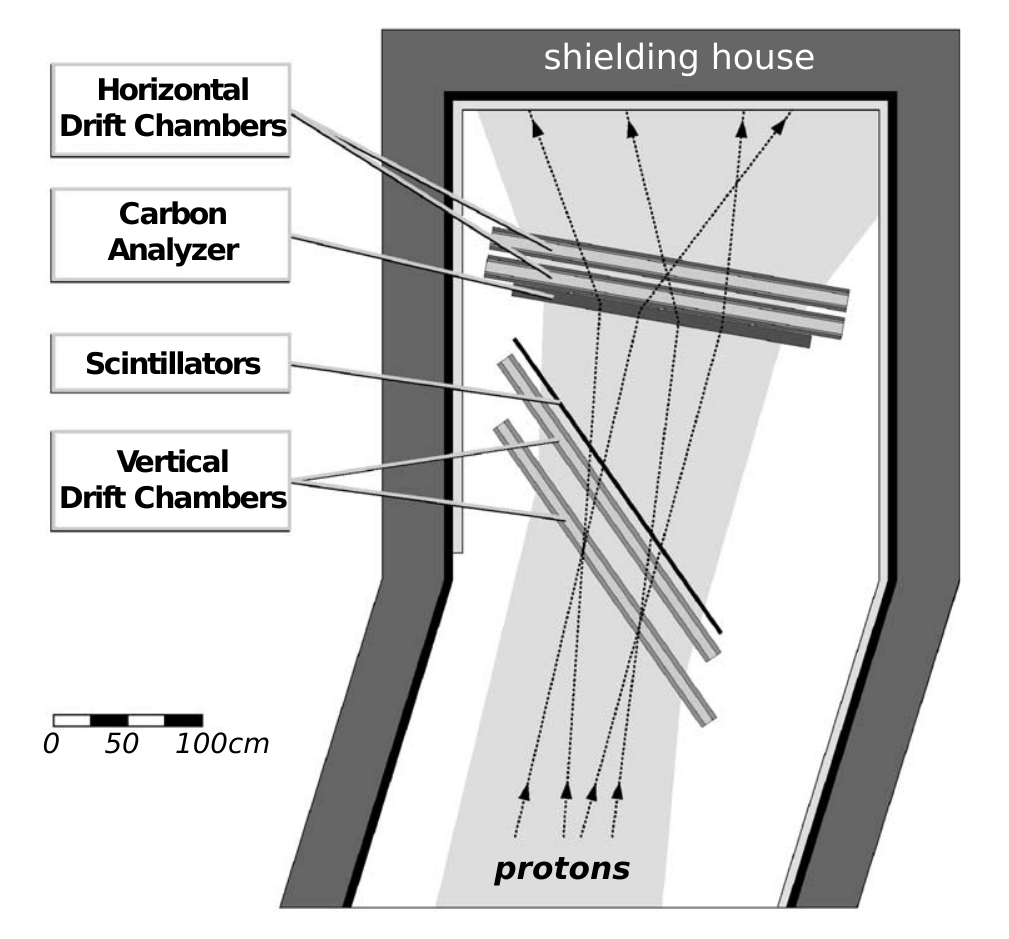}
\caption{
Sideview of the detector system of the proton spectrometer. The standard detector system (VDCs and scintillators) is supplemented by the FPP, which consists of a carbon analyzer and two double-planes of HDCs for proton tracking after scattering in the carbon analyzer. 
Some proton trajectories are indicated within the region of acceptance (shaded area).  Adapted from \cite{Pospischil:2000pu}.
}
\label{fig:fpp}
\end{figure}

 The proton spectrometer was equipped with a focal-plane-polarimeter (FPP) consisting of a 7 cm thick carbon analyzer \cite{a1aparatus,Pospischil:2000pu} and horizontal drift chambers (HDC) to measure the secondary scattering of the proton in the analyzer, as shown in Fig.~\ref{fig:fpp}.  
 The spin-dependent scattering of the polarized proton by the carbon analyzer enables the determination of the proton's transverse polarization components at the focal plane \cite{Pospischil:2000pu}. The polarization-transfer components at the \textit{reaction} point were obtained by transforming the measured components using the known spin precession in the magnetic field of the spectrometer. 

\section{Measured reaction and kinematics\label{sec:kin}} 

\begin{figure}[ht]
\includegraphics[width=\columnwidth]{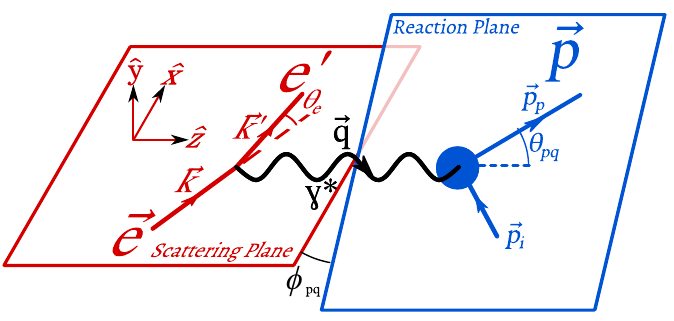}
\caption{
Kinematics of the measured reaction.
}
\label{fig:kinematics_diagram}
\end{figure}

 The kinematics of the measured reaction are shown in Fig.~\ref{fig:kinematics_diagram}.  The electron's initial and final  momenta are $\vec k$ and $\vec k'$ respectively, and they define the scattering plane of the reaction.  The proton's initial momentum introduces, in addition, the reaction plane defined by the momentum transfer $\vec q = \vec k-\vec k\,'$ and the exiting proton's momentum $\vec p\,'$.  The angle between the scattering plane and the reaction plane is denoted by $\phi_{pq}$.  
 
 Following the convention of \cite{Strauch}, we express the components of the polarization transfer $\vec P\,'$ in the scattering-plane coordinate system, where $\hat z$ is along the direction of the momentum transfer $\vec q$, $\hat y$ is along the direction of $\vec k \times \vec k\,'$, and $\hat x = \hat y\times\hat z$, forming a right-handed coordinate system.  

The missing momentum  $\vec p_{\rm miss}= \vec q - \vec p\,'$ is the recoil
momentum of the residual nucleus. Neglecting FSI, 
$-\vec p_\mathrm{miss}$ is equal to the initial momentum of the emitted proton.
We conventionally define positive and negative signs for $p_{\rm miss}$ as the sign of $\vec p_{\rm miss}\cdot\vec q$.

Our data were taken at two kinematic settings, A and B, covering different ranges in $p_{\rm miss}$ at different  invariant four-momentum transfers, $Q^2 = - q^{2}$.  Setting A was centered near $p_{\rm miss}=0$, at $Q^2 = 0.40~({\rm GeV}\!/\!c)^2$.  Setting B covered a region of large negative $p_{\rm miss}$, at $Q^2 = 0.18~({\rm GeV}\!/\!c)^2$.  The details of these kinematic settings are given in Table \ref{tab:kinematics}.

\begin{table}[ht!]
\caption{
The kinematic settings in the $^{12}$C($\vec{e},e'\vec p$) experiment. The angles and momenta represent the central values for the two spectrometers: $p_p$ and $\theta_p$ ($p_e$ and $\theta_e$) are the knocked out proton (scattered electron) momentum and scattering angles, respectively.
}
\begin{center}
\begin{tabular}{lllll}
\hline
& & \multicolumn{2}{l}{Kinematic Setup}  & \\ \cline{3-5}
 & &  A & B \\
\hline
$E_{\rm beam}$ & [MeV] & 600 & 600 \\
$Q^2$ & [\GeVsq] & 0.40  & 0.18\\
$p_{\rm miss}$ & [MeV] & $-$130 to 100  &$-$250 to $-$100 \\
$p_e$ & [MeV/$c$] & 385 & 368 \\
$\theta_e$ & [deg] & 82.4 & 52.9 \\
$p_p$ & [MeV/$c$] & 668 & 665 \\
$\theta_{p}$ & [deg] & $-$34.7  & $-$37.8 \\
\multicolumn{2}{l}{\small{\# of events after cuts}} & 1.7 M &  1.1 M \\
\hline
\end{tabular}
\end{center}
\label{tab:kinematics}
\end{table}

\begin{figure}
\includegraphics[width=\columnwidth]{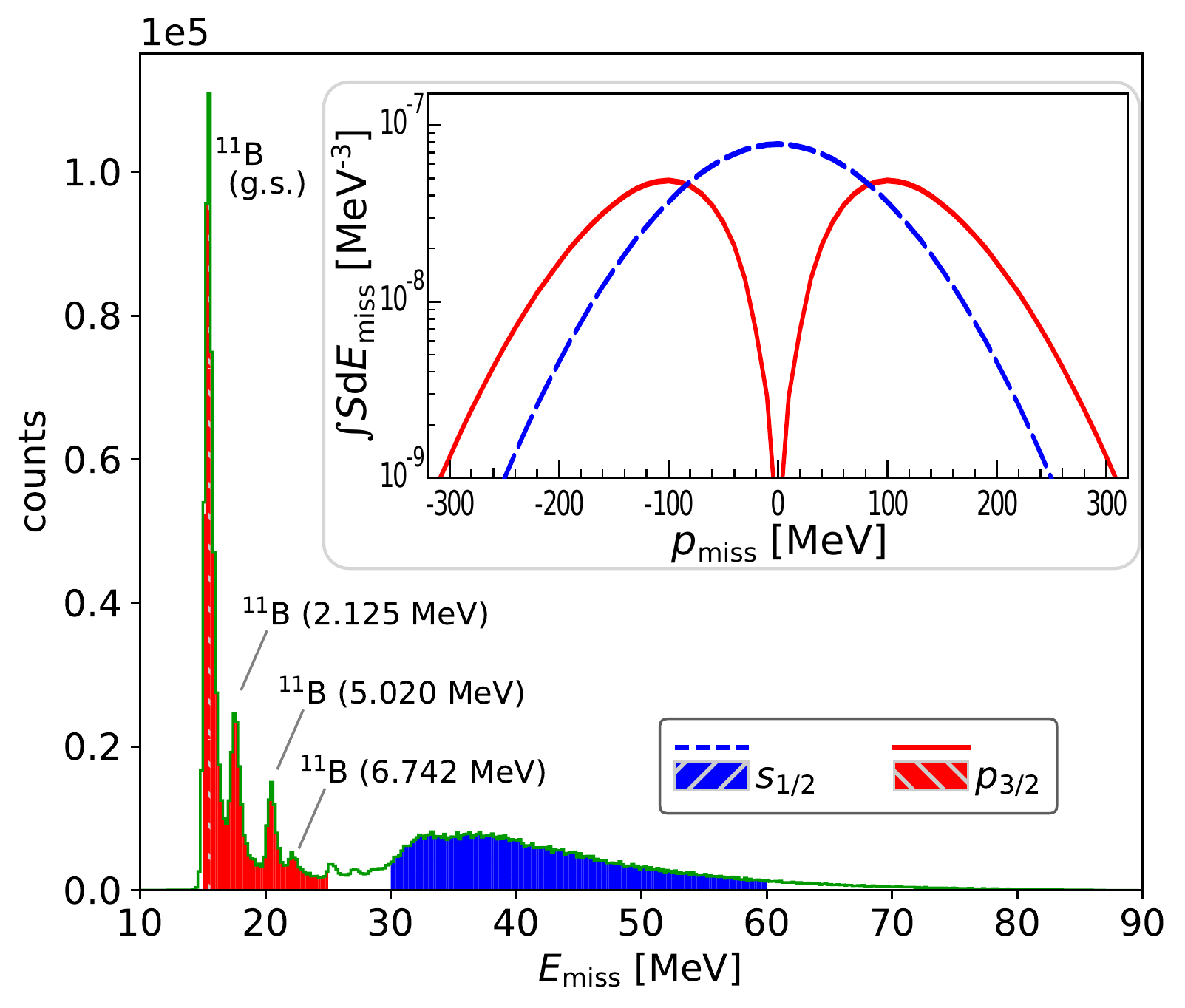}
\caption{
The measured proton missing-energy spectrum for $^{12}{\rm C}(\vec e, e' \vec p)$ (data shown are for Setup A). The distinct peaks correspond to removal of $p_{3/2}$ -shell protons in $^{12}$C resulting in $^{11}$B ground
state and excited states as noted. The $E_{\rm miss}$ ranges considered in the analysis for $p_{3/2}$ and $s_{1/2}$ protons are marked in red and blue, respectively (color online). The inset shows the momentum distribution predictions of the independent-particle shell model for $p_{3/2}$ and $s_{1/2}$ protons in $^{12}$C, adapted from \cite{Dutta}.  
}
\label{fig:emiss_pmiss}
\end{figure}

Following \cite{Dutta}, we distinguish between protons knockout from the $s$- and $p$-shells using cuts on the missing energy, $E_{\rm miss}$ in the reaction, defined as:

\begin{equation}
E_{\rm miss} \equiv \omega - T_p - T_{^{11}\rm B},
\end{equation}
 where $\omega=k^0-k'^{\,0}$ is the energy transfer, $T_p$ is the measured kinetic energy of the outgoing proton, and $T_{^{11}\rm B}$ is the calculated kinetic energy of the recoiling $^{11}$B nucleus.
 The measured $E_{\rm miss}$ spectrum is shown in Fig.~\ref{fig:emiss_pmiss}.  
    For the $s$-shell sample, we used the cut 30 MeV $< E_{\rm miss} < 60$ MeV,  while for the $p$-shell sample, we used 15 MeV $< E_{\rm miss} < 25$ MeV \cite{ceepLet}.   Also shown in Fig.~\ref{fig:emiss_pmiss} are the $p_{\rm miss}$ spectra for the two shells.
 
 When protons are knocked out from the $p$-shell of the carbon nucleus, the $A-1$ system is left in a discrete state (either the ground state or one of the excited states of $^{11}\rm B$).  However, knockout from the $s$-shell leaves the residual $A-1$ system unbound, leading to a continuum of residual states.  
 The wave function at $p=0$ has a minimum for the $p$-state and a maximum for the $s$-state (see inset of Fig.~\ref{fig:emiss_pmiss}).

An important quantity characterizing the proton
prior to its knock-out is its ``off-shellness''. We quantify this using the
virtuality, $\nu$, a variable defined as  \cite{deep2012PLB}

\begin{equation}
\label{eq:virt}
\nu \equiv \left(M_Ac-\sqrt{M^2_{A-1}c^2+p_{\rm miss}^2}\right)^2-p_{\rm miss}^2-M^2_p c^2,
\end{equation}
where $M_A$ is the mass of the target nucleus, $M_{A-1}\equiv\sqrt{\left(\omega-E_p +M_Ac\right)^2-p_{\rm miss}^2}$ is the mass of the residual nucleus (not necessarily in its ground state) determined event by event, and $E_p$ is the total energy of the outgoing proton.  We note that the virtuality depends not only on $p_{\rm miss}$, but also on $M_{A-1}$.   The minimum value of $|\nu|$ (for a given target nucleus) is  

\begin{equation}
\label{eq:minVirt}
|\nu|_{\rm min} = \left(M_p^2-\left(M_{A}-M_{A-1}^{\rm g.s.}\right)^2\right)c^2,
\end{equation}
where  $M_{A-1}^{\rm g.s.}$ is the ground-state mass of the the residual nucleus (in this case, $^{11}\rm B$).  For the $^{12}{\rm C}(\vec e,e'\vec p)^{11}\rm B$ reaction, $|\nu|_{\rm min}$ = 0.0297 $({\rm GeV}\!/\!c)^2$.  Protons knocked out from the $s$-shell are generally further off-shell than those in the $p$-shell (even in events at the same $p_{\rm miss}$, due to their larger $E_{\rm miss}$).

\section{Data analysis}
\label{sec:measurement}
\subsection{Event reconstruction and selection}
\newcommand{\code}{\texttt}
For the event reconstruction, we used the \code{Cola++} reconstruction framework \cite{Cola}, developed by the A1 collaboration.  The  reconstructed angles and positions of the tracks at the VDC plane are used 
in a fit to find the initial momentum, angle and vertex position of each particle.  A multivariate polynomial fit is then used to reconstruct the proton's  spin-transfer matrix (STM) for each event.

Software cuts were then applied to the data.  We require the time-coincidence between the scintillators of the two spectrometers to be within a $\pm$5 ns window, and that the signal in the \v Cerenkov detector be above the electron-detection threshold.  We include tracking cuts which ensure good tracking in the spectrometers and require both the electron and the proton originate from the carbon target.  Further tracking cuts require the proton's trajectory to be within the part of the spectrometer where the precession of the proton's spin is well known.

In the FPP, we removed events with (spin-independent) Coulomb-scattering in the carbon analyzer by selecting only events where the proton scattered by more than 8$\degree$.
We also required the fitted position of the scattering point in the FPP (determined by the HDC and the extrapolated trajectory from the VDC) to be consistent with the actual position of the analyzer.

\subsection{Polarization fitting}
There are two types of polarization observables that can be obtained in this experimental setup: the induced polarization, $\vec P$, (which is the proton polarization for an unpolarized electron beam) and the polarization transfer, $\vec P\,'$, (which determines the beam polarization-dependent part of the proton polarization).  The total outgoing proton polarization is related to these observables via:

\begin{equation}
\vec P_{p,\rm tot} = \vec P + hP_e\vec P\,',
\end{equation}
where $h$ and $P_e$ are the helicity and polarization of the electron beam.

We perform a fit to obtain $\vec P$ and $\vec P\,'$ by maximizing the log-likelihood

\begin{equation}
\log \mathcal{L} = \sum\limits_k \log (1+\vec P_{p,\rm tot} \cdot \vec\lambda_k),
\end{equation}
where the $\vec \lambda_k$ for each event are given by

\begin{equation}
\vec \lambda_k = a \mathbf{S}^{-1} \left(\begin{array}{c}-\sin{\phi_{\rm FPP}} \\\cos{\phi_{\rm FPP}} \\0\end{array}\right),
\end{equation}
and $\phi_{\rm FPP}$ is the measured azimuthal scattering angle in the FPP.~  $\mathbf{S}$ is the calculated spin-transfer matrix for the event and $a$ is the analyzing power of the event (as determined in \cite{AprileGiboni:1984pb,Mcnaughton:1986ks}).  

Three of these components ($P_x$, $P'_y$ and $P_z$) are expected to be very small for individual events and antisymmetric in their dependence on $\phi_{pq}$ \cite{Giusti:1989ww}.  Hence, they would average to zero when considering event samples that have symmetric distributions in $\phi_{pq}$, as is nearly the case in our dataset.  
Therefore, in order to improve the stability of our fit, we fix these parameters to zero\footnote{We have also performed the fit with all six parameters included, and have found that this does not strongly affect the outcome of the fit except in bins with poor statistics.}, leaving us with

\begin{equation}
\vec P_{p,\rm tot} = \left(\begin{array}{c}hP_eP'_x\\P_y\\hP_eP'_z\end{array}\right).
\end{equation}

\begin{figure*}[t!]
\subfigure[\normalsize\ Polarization-transfer components and their ratio versus missing momentum.]{
\includegraphics[width=0.979\columnwidth]{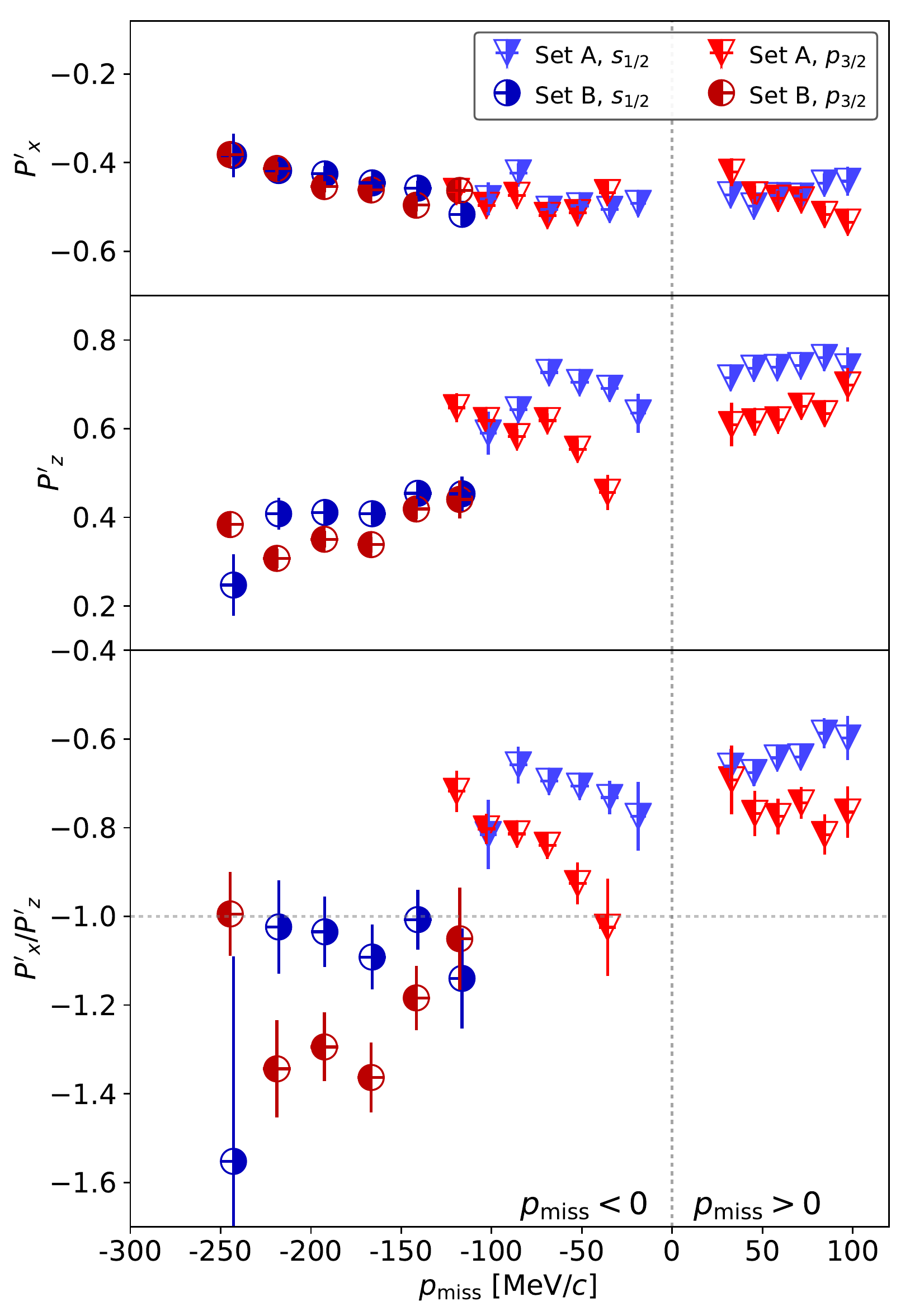}
\label{subfig:raw_vs_pmiss}
}
\subfigure[\normalsize\ Polarization-transfer components and their ratio versus virtuality.]{
\includegraphics[width=1.021\columnwidth]{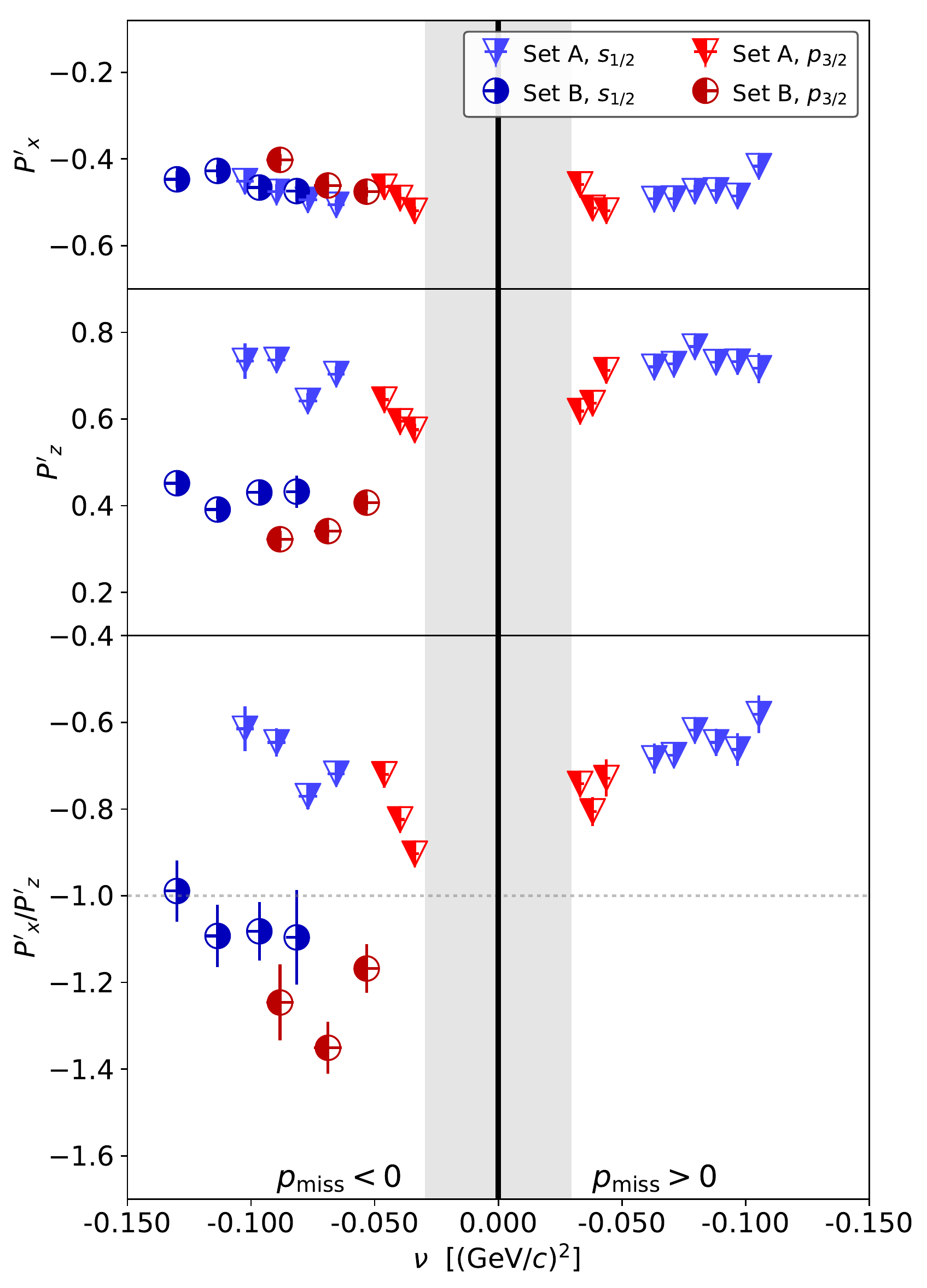}
\label{subfig:raw_vs_virt}
}
\caption{The measured polarization-transfer components, $P'_x$ (top), $P'_z$ (middle), and their ratio, $P'_x/P'_z$ (bottom).  These are plotted versus the missing-momentum \protect\subref{subfig:raw_vs_pmiss} and virtuality \protect\subref{subfig:raw_vs_virt}.  In the virtuality plot \protect\subref{subfig:raw_vs_virt}, the data with positive (negative) $p_{\rm miss}$ are shown separately.  The grey band shows the kinematically forbidden region at $|\nu|<$ 0.0297 $({\rm GeV}\!/\!c)^2$  (see Eq.~\ref{eq:minVirt}).   The uncertainties shown are statistical only.  Systematic errors are discussed in Section \ref{sec:systematics}.   Triangles (circles) refer to kinematic Setting A (B).~ Symbols that are open on the left (right) side refer to $s$- ($p$-) shell removals, and are colored blue (red) online.  The legend is common to all panels in the figure.
}
\label{fig:raw_vs_pmiss_and_virt}
\end{figure*}

We partitioned both the $s$- and $p$-shell knockout data into bins by $p_{\rm miss}$, and performed the above procedure on each bin separately, the results of which are shown in Fig.~\ref{subfig:raw_vs_pmiss}.  Likewise, we binned the data by the virtuality, and show the results in Fig.~\ref{subfig:raw_vs_virt}.  
We present the results binned by both variables to show how the polarization may vary with the protons' motion ($p_{\rm miss}$ bins) and also their off-shellness (virtuality bins).  This avoids conflating effects related to one variable with those related to the other, as the two are correlated.  

\subsection{Systematic uncertainties in the measurements}
\label{sec:systematics}
The systematic
errors in these measurements are due to a few sources, which are presented in Table \ref{tab:systematics}.
The largest contribution to the uncertainty in the polarization
components $P'_x$ and $P'_z$ is due to the beam polarization, which was determined with an estimated accuracy of 2\%.  These components are also sensitive to the analyzing
power of the carbon secondary scatterer, which in this kinematic region is known to about 1\% \cite{Pospischil:2000pu,AprileGiboni:1984pb,Mcnaughton:1986ks}. 
However, the ratio $P'_x/P'_z$ is independent of both the beam polarization and the analyzing power. 

The uncertainties in the  beam energy and the central kinematics of the spectrometers in each dataset affect the basis vectors that define the scattering-plane coordinate system, as well as which bin an event goes into.  
The alignment between the HDC coordinate system and the tracks extrapolated from the VDC to the HDC plane  also affected the polarization measurement, since these measurements depend on the distribution of the angles of the secondary scattering.

These three sources of uncertainty (beam energy, central kinematics, and detector alignment) were determined in the following manner.  We modified each of the variables one-by-one by their uncertainty values, and repeated the analysis, and then determined how much this affected the extracted polarizations.  The errors from each source were then added in quadrature.  
  
  \begin{table}[ht]
\caption{
Systematic uncertainties of $P'_x$, $P'_z$ and $P'_x/P'_z$.  All values are in percent.  See text for details.}

\begin{tabular}{lrrr}
\hline
 & $dP'_x/P'_x$  & $dP'_z/P'_z$ & $\frac{d[P'_x/P'_z]}{P'_x/P'_z}$ \\
\hline
Beam pol. & 2.0  & 2.0 & 0.0 \\
Analyzing power & 1.0  & 1.0 & 0.0 \\
Beam energy & 0.2  & 0.6 & 0.8 \\
Central kinematics & 0.6 & 0.8 & 0.9 \\
Alignment & $<$0.1 & 0.1 & 0.1 \\
Software cuts & 1.7  & 2.1 & 1.9 \\
Precession (STM fit) & 0.3  & 0.3 & 0.3 \\
Precession (trajectory) & 0.2 & 0.3 & $<$0.1 \\
\hline
Total & 2.9  & 3.3 & 2.3 \\
\hline
\end{tabular}
\label{tab:systematics}
\end{table}
  
The systematics due to software cuts were studied by reanalyzing the data with each cut slightly tighter than the actual value that is used in the final analysis and taking the average of the effects of the modified cut over all of the bins.  The values in the row labeled ``software cuts'' in Table \ref{tab:systematics} are the quadratic sums of the effects from each of the different cuts.    
  
The uncertainty in the
spin-precession evaluation was estimated by comparing the STM calculated internally by the \code{Cola++} software in the event reconstruction (which uses a polynomial fit) and a more precise (but much slower) calculation using the \code{QSPIN} program \cite{Pospischil:2000pu}.  The fit was able to reproduce the spin-precession with an accuracy of 0.3\% \cite{Pospischil:2000pu}.  

Furthermore, the finite resolution of the proton's trajectory parameters, especially the vertex position, adds additional systematic uncertainty to the precession.  To begin determining this part of the systematic uncertainty, we used \code{QSPIN} to calculate the spin-transfer matrix, $\mathbf{S}_{\rm ref}$, for a reference trajectory.  We then produced 100 other trajectories 
with normally distributed 
variations in each parameter, where the standard
 deviation of each parameter equals the resolution of that parameter, to produce precession matrices $\mathbf{S}_i$.  The resulting uncertainty on the measured polarization due to the trajectory is then 

\begin{equation}
\delta \vec P =  (\mathbf{I}_3 - {\rm{avg}}_i[\mathbf{S}_i] \mathbf{S}_{\rm ref}^{-1})\vec P,
\end{equation}
where $\mathbf{I}_3$ is the $3\times3$ identity matrix and ${\rm avg}_i [\mathbf{S}_i]$ represents the average of the matrices $\mathbf{S}_i$.  The average is performed because the measured polarization is calculated using an ensemble of trajectories, rather than a single trajectory.

The total estimated systematic
uncertainties are calculated by adding the effects of each of the  individual sources in quadrature.  The systematic uncertainties for the individual transfer components $P'_x$ and $P'_z$ are  $\sim$3\% and the uncertainty for their ratio is $\sim$2\%.  These are consistent with estimates of the systematic uncertainties from previous polarization-transfer measurements on the deuteron using the same experimental setup and comparable kinematic settings \cite{deep2012PLB,ceepLet,deepCompPLB,posPmissPLB}.

\section{Interpretation of results\label{sec:interp}}

\subsection{General observations}
As shown in Fig.~\ref{fig:raw_vs_pmiss_and_virt}, the $P'_x$ components are less sensitive to the different kinematic variables, $p_{\rm miss}$ and $\nu$, than $P'_z$.  We find that $P'_x$ is nearly identical for $s$- and $p$- shells at the same $p_{\rm miss}$.    The $P'_z$ component for $p$-shell knockout dips down at small negative $p_{\rm miss}$, while this does not
seem to be the case in $s$-shell knockout.   

We observe that where the two kinematic settings overlap in virtuality, there is no regularity in $P'_z$, and that $P'_z$ is much larger at Setting A than at Setting B.

In order to further interpret our polarization-transfer data, we compare them to dedicated calculations of $^{12}{\rm C}(\vec e,e'\vec p)$ described in Section \ref{sec:calc} below.  

\subsection{Calculations of the polarization transfer for $^{12}$C}
\label{sec:calc}

We compared the measured polarization transfer to relativistic
distorted-wave impulse approximation (RDWIA) calculations \cite{Meucci:2001qc} 
where the FSI between the outgoing nucleon and the residual nucleus are described
by a phenomenological relativistic optical potential.  

In the calculations, the so-called ``democratic" optical potential \cite{PhysRevC.80.034605} has been used, which has been obtained using a global fit to over 200 datasets using elastic proton-nucleus scattering over a broad range of nuclei from helium to lead.  The differences in the calculated polarization transfer due to the choice of the optical potential were estimated at about 2-4\%.  The evaluation was done by comparing the
results obtained with the democratic and the energy-dependent and
atomic-number-independent (EDAI) relativistic optical potential \cite{Cooper:1993nx},
which is a single-nucleus parameterization, constructed to reproduce
elastic proton-scattering data just on $^{12}$C.

The relativistic bound-state wave functions used in the calculations have been
obtained by solving the relativistic
Hartree-Bogoliubov Equations using the program \code{ADFX} \cite{Poschl:1997hg}. The model is applied in the
mean-field approximation to the description of ground-state properties
of spherical nuclei, using a Lagrangian containing the $\sigma$, $\omega$ and
$\rho$ mesons and the photon field \cite{Serot:1984ey,Serot:1997xg,Ring:1996qi,Lalazissis:1996rd}. Moreover,
finite-range interactions are included to describe pairing
correlations and the coupling to particle continuum states.
The Lagrangian parameters are usually obtained by a fitting procedure
to some bulk properties of a set of nuclei. The wave-functions used in our calculations were obtained
with the NL-SH parametrization \cite{SHARMA1993377}. The results of the calculations using different parameterizations 
differ by about 0.5-0.8\%.

The Coulomb
distortion of the electron wave functions is considered using the
effective-momentum approximation.
Our calculation uses the parameterization of the free-proton FFs from \cite{Bernauer}, which are known to within 0.5\% in the kinematic region of our experiment.

In coplanar kinematics, a set of 8 structure functions contribute to the polarization transfer \cite{Giusti:1989ww,Boffi:1996ikg}.
In \textit{non}-coplanar kinematics, an additional structure function, $\bar h'^N_{01}$, contributes to $P'_x$, but not $P'_z$.

The RDWIA program \cite{Meucci:2001qc} was written to perform calculations only in the
coplanar kinematics of the usual $A(\vec e,e'\vec p)$ experiments. Therefore, it
calculates only the structure functions that contribute in coplanar
kinematics and not $\bar h'^N_{01}$.
We performed calculations in non-coplanar kinematics using the approximation

\begin{equation}
\label{eq:approx1}
\bar h'^N_{01} = -\bar h'^S_{01},
\end{equation}
where $\bar h'^S_{01}$ is one of the structure functions calculated in \cite{Meucci:2001qc}.  Eq.~\ref{eq:approx1} is exactly true in parallel kinematics ($\theta_{pq} = 0$) \cite{Giusti:1989ww}, and we assume that it provides a valid approximation at small $\theta_{pq}$, which dominate the kinematics of our data.

\begin{figure*}[t!]
\begin{center}
\subfigure[\normalsize{\ $s$-shell knockout}]{
\includegraphics[width=\columnwidth]{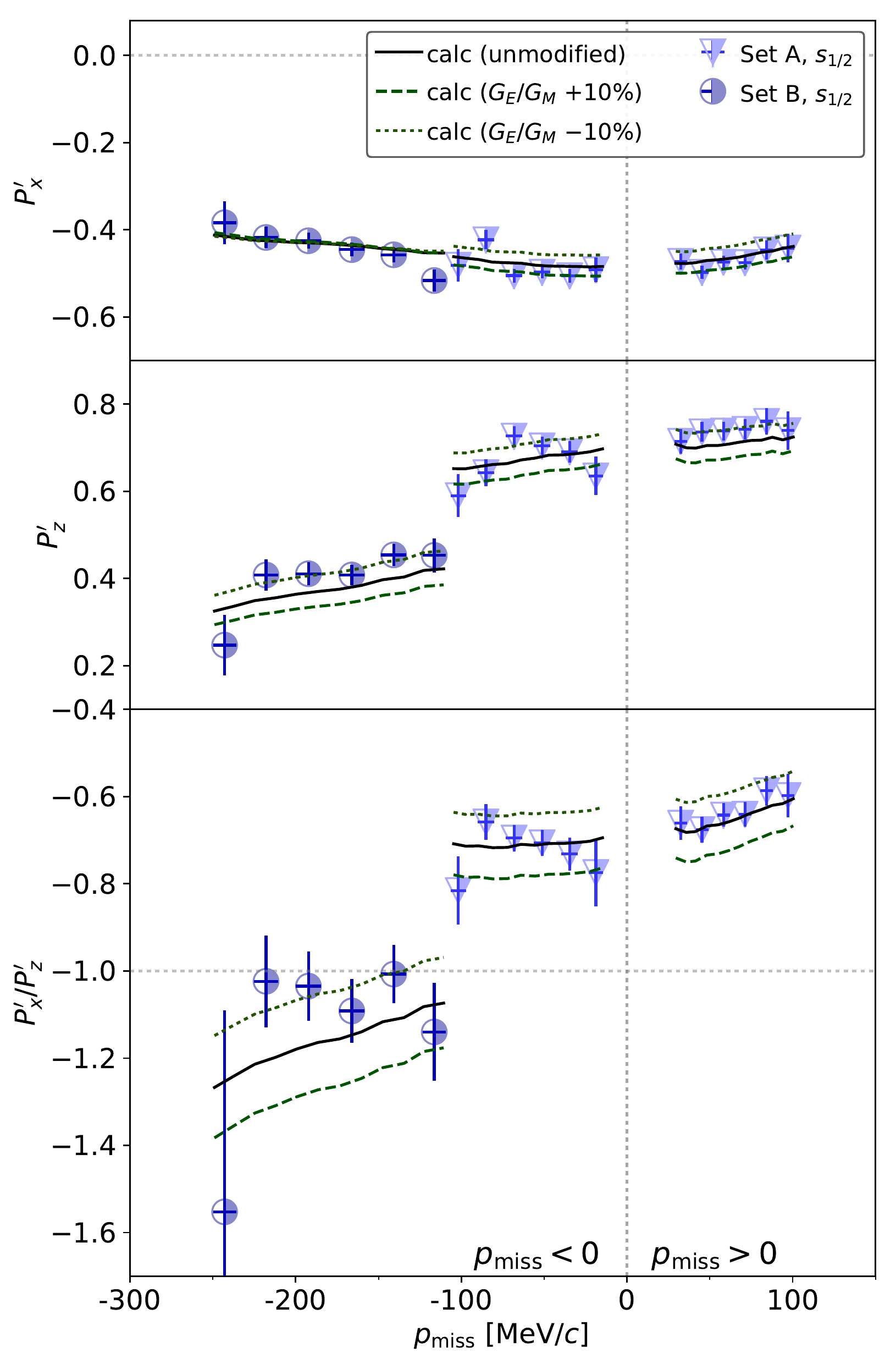}
\label{fig:wf1_dem_s}
}
\subfigure[\normalsize{\ $p$-shell knockout}]{
\includegraphics[width=\columnwidth]{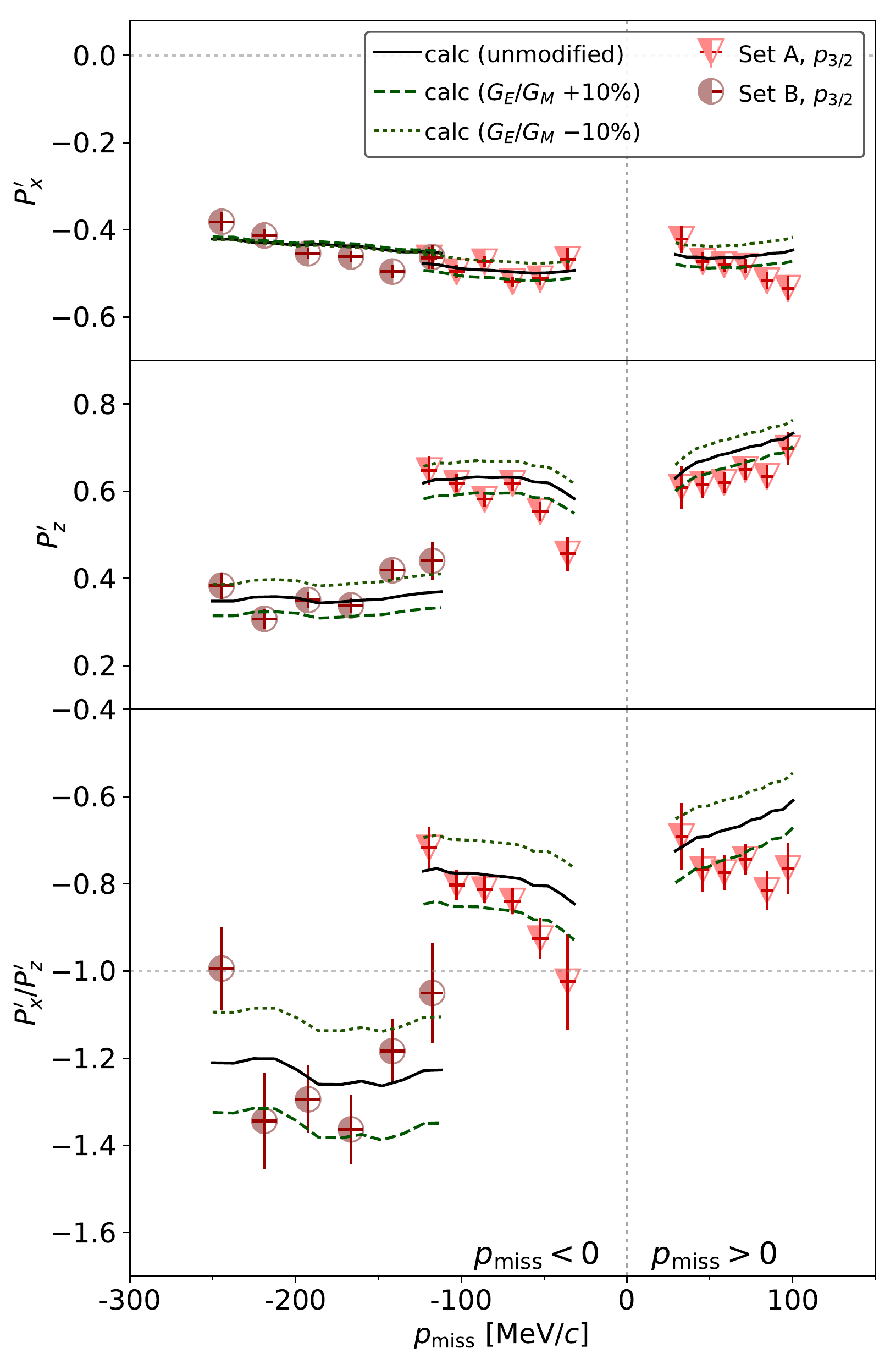}
\label{fig:wf1_dem_p}
}
\caption{
DWIA calculations of the polarization observables, with unmodified FFs (solid curves), and with $G_E/G_M$ scaled by +10\% ($-$10\%) shown as long-dashed (dotted) curves.  These calculation curves are overlaid over the measured values from Fig.~\ref{subfig:raw_vs_pmiss}.  
These are shown separately for $s$-shell knockout \protect\subref{fig:wf1_dem_s} (blue online) and $p$-shell knockout  \protect\subref{fig:wf1_dem_p} (red online). 
}
\label{fig:wf1_dem}
\end{center}
\end{figure*}

We averaged the polarizations calculated for the kinematics of a sample of events in each $p_{\rm miss}$ bin and show the results in Fig.~\ref{fig:wf1_dem} as solid curves, compared to our data.  
In order to examine the sensitivity of the calculations to the proton FFs, we show as long-dashed (dotted) curves the results obtained with
$G_E/G_M$ rescaled by +($-$) 10\%.

For Setting B, the curves for $P'_x$ at $\pm$10\% are very close to the calculations for unmodified FFs, indicating that $P'_x$ is not sensitive to the FF ratios in this region.  This makes the $P'_x$ at large $p_{\rm miss}$ useful for testing the agreement between the calculations and the data.  By contrast, $P'_z$ is strongly sensitive to the FF scaling.  At Setting A, on the other hand, both $P'_x$ and $P'_z$ are sensitive to the scaling of the FF ratio.  

For the $s$-shell knockout at both kinematic settings, there is excellent agreement between the  calculations and the data, except in $P'_z$ at large negative $p_{\rm miss}$ (Setting B), where they differ by about 10\%.  One may consider the possibility of scaling $G_E/G_M$ in order to obtain a better fit to the data.  

For the $p$-shell knockout, the calculations and the data are in decent agreement with each other, but the agreement is not as good as for the $s$-shell.  Specifically, at Setting A, $P'_z$ is overestimated by the calculations, while the magnitude of $P'_x$ appears to be somewhat underestimated by the calculations.   

Before quantifying the agreement between the calculations and the data,  we discuss in Sec.~\ref{sec:calcsysunc}  the uncertainties and limitations of the calculations.   

\subsection{Uncertainties of the calculations}
\label{sec:calcsysunc}

We have adopted a model (RDWIA) which is based on some
assumptions and approximations which may affect the comparison with the
data.  Within the model there are uncertainties due to
the choice of the different ingredients (bound-state wave functions, optical
potentials, and proton form factors) that are adopted in the calculations.
We evaluated the combined contribution to the uncertainty in the calculation due to the parameterizations of these ingredients to be 2.2\%, 3.8\% and 4.0\% for $P'_x$, $P'_z$, and  $P'_x/P'_z$, respectively.  These were obtained by adding the estimated contributions, as discussed in Section~\ref{sec:calc}, from those three sources in quadrature.   By far the largest contribution to this uncertainty comes from the choice of the optical potential.  

In the calculations for the $p$-shell knockout, it is assumed that the $^{11}$B residual nucleus is left in its ground state.  The data include also the excited states, as shown in the measured missing-energy spectrum of Fig.~\ref{fig:emiss_pmiss}.  However, the majority of the $p$-shell contribution comes from the ground state and we have checked that the excited states do not strongly affect the data for $p$-shell knockout.  Furthermore, the wave-function of the $p$-shell has a minimum at $p=0$, possibly reducing the numerical accuracy of the calculations at low $p_{\rm miss}$ for this shell.  The $s$-shell wave-function
does not have this problem, but the fact that in this case the  $A-1$
system is in the continuum, while the DWIA model is in principle
devised for discrete states, introduces an additional (although presumably
reasonable) approximation.
Finally, we note that in the calculations a relativistic
mean-field approximation has been assumed. A more sophisticated model
for nuclear structure 
may impact the results and improve the agreement with the experimental data.

\subsection{Quantifying the agreement of the calculations with the data}
To quantify the agreement between the calculations and the data, we use a bin-by-bin comparison, employing the $\chi^2$ criterion:

\begin{equation}
\label{eq:chi2}
\chi^2 = \sum\limits_{\substack{\alpha\,\in\,\{x,z\}\\ i\,\in\,\rm bins\,}}\frac{(P'^{\,\rm meas}_{\alpha,i}-P'^{\,\rm calc}_{\alpha,i})^2}{(dP'^{\,\rm meas}_{\alpha,i})^2+(dP'^{\,\rm calc}_{\alpha,i})^2},
\end{equation}
where $P'^{\,\rm meas}_{\alpha,i}$ and $P'^{\,\rm calc}_{\alpha,i}$ are the measured and calculated values of the polarization transfer for a given axis, $\alpha$ ($x$ and $z$), in the $i^{\rm th}$ bin.  The uncertainty in the measurement, $dP'^{\,\rm meas}_{\alpha,i}$, includes the statistical error of the fit and the systematic uncertainty\footnote{about 3\% for both $P'_x$ and $P'_z$, see Table \ref{tab:systematics}.} added in quadrature.  For $dP'^{\,\rm calc}_{\alpha,i}$, we only included the errors of the calculation due to the parameterizations of the optical potential, wave-function, and free-proton FFs (see Sec.~\ref{sec:calcsysunc}), since the impacts of other effects on the calculations are undetermined.   The $\chi^2$ values are then converted to $p$-values, denoted by $p_{\rm val}[\chi^2]$, where the number of degrees of freedom is twice the number of bins, as there is a measurement of $P'_x$ and $P'_z$ for each bin.  

Scaling the FF ratio $G_E/G_M$ has been suggested above in order to obtain a better fit to the data.  In order to determine if this scaling improves the agreement between the calculations and the data, we varied the value of

\begin{equation}
 a_{\rm mod} \equiv \left(\frac{G_E}{G_M}\right)^{\rm bound}\left/\middle(\frac{G_E}{G_M}\right)^{\rm free} -1,
 \end{equation} 
and then reevaluated $\chi^2$ and the $p$-values.  
We determined the optimal values of $a_{\rm mod}$ which minimized the $\chi^2$ separately for the $s$- and $p$-shells in each kinematic setting.
 The uncertainty on the fitted values of $a_{\rm mod}$ are given by 

 \begin{equation}
 \delta a_{\rm mod} = \left[\frac{\partial^2}{\partial a_{\rm mod}^2}[\chi^2]\right]^{-1/2},
 \end{equation}
 evaluated where $\chi^2$ is at its minimum.  
 The results of these comparisons are given in Table \ref{tab:chi2}.

  \begin{table}[h]
\caption{The $p$-values for the fits with unscaled and optimally scaled FFs. We also include the $a_{\rm mod}$ for the optimal scaling.  The number of degrees of freedom for the comparison with scaled FFs is one fewer than for those with unscaled FFs.} 
\setlength{\tabcolsep}{4pt}
\begin{tabular}{c|c|cc|r|cc}
\multicolumn{2}{c|}{shell/} & \multicolumn{2}{c|}{w/ unmod. FFs} & \multicolumn{1}{c|}{$a_{\rm mod}$} & \multicolumn{2}{c}{w/ mod. FFs}\\
\multicolumn{2}{c|}{setting} & $n_{\rm dof}$ & $p_{\rm val}[\chi^2]$ & \multicolumn{1}{c|}{(\%)}& $n_{\rm dof}$ & $p_{\rm val}[\chi^2]$ \\
\hline
\multirow{2}{*}{$s_{1/2}$} & A &24 & 0.91 & $0.3\pm2.5$ & 23 & 0.89 \\
 & B &12 & 0.29 & $-8.6\pm3.8$ & 11 & 0.60 \\
\hline
\multirow{2}{*}{$p_{3/2}$} & A &24 & 0.017 & $10.4\pm2.6$ & 23 & 0.37 \\
& B &12 & 0.049 & $-2.5\pm3.1$ & 11 & 0.04 \\
\hline
\end{tabular}
\label{tab:chi2}
\end{table}

 This comparison shows that the $s$-shell calculations at both kinematic settings agree with the data, even without scaling the FFs.  
 At Setting B the optimized scaling $a_{\rm mod}=-8.6\%$ improves the agreement for
$P'_z$, while at Setting A there is no need of FF scaling.

 The $p$-shell calculations, without and even with FF scaling, are 
generally in worse agreement with the data than those for the $s$-shell.
However, as shown in Table \ref{tab:chi2}, a scaling of $a_{\rm mod}=10.4\%$ at Setting A
greatly improves the agreement with the data. This is an indication of either genuine modifications to the FFs, or of other possible effects which mimic a modification to the FFs.   At Setting B, no FF scaling improves the agreement with the data.

  \section{Conclusions}

Measurements of the transverse and longitudinal components of the polarization transfer in quasi-free $^{12}{\rm C}(\vec e,e'\vec p)$ reaction have been presented and compared to RDWIA calculations.  The comparison gives an overall good agreement, but some discrepancies are observed
in $s$-shell knockout at large negative $p_{\rm miss}$ and in $p$-shell knockout.

For $s$-shell knockout, both $P'_x$  and $P'_z$ are in very good agreement with the data at low $p_{\rm miss}$.  $P'_x$ is in good agreement with the data also at high $p_{\rm miss}$.  
The only notable discrepancy for the $s$-shell is that the calculations  underestimate $P'_z$ by about 10\% at large negative $p_{\rm miss}$.  
The $s$-shell knockout at large $p_{\rm miss}$ is a region of particular interest when searching for medium modifications to the form factors \cite{PhysRevC.87.028202}, because it has both the largest virtuality and the largest nuclear density in our dataset. 
We note that in this region, $P'_x$ is insensitive to $G_E/G_M$, and therefore the agreement between the calculated and measured $P'_x$ gives further credibility to the calculations.  
 Modifying the form-factor ratio in the calculations  by $-8.6\%$ in this region improves their agreement with the data (by scaling $P'_z$ while keeping $P'_x$ unchanged).

For the $p$-shell knockout, the agreement with the data is not as good as for the $s$-shell.   At low $p_{\rm miss}$, scaling the form-factor ratio by +10.4\% greatly improves the agreement with the data, which however, remains worse than for the $s$-shell.  At high $p_{\rm miss}$, scaling the form-factor ratio does not improve the agreement with the data. Further theoretical work is required to understand the source of the discrepancies for $p$-shell knockout.

No global scaling to $G_E/G_M$, common to all kinematic regions in our dataset, would solve all the discrepancies.
Scaling the form-factor ratio differently for each kinematic region leads to a good agreement between the data and calculations.  We note, however, that form-factor modification is not the only possible solution.  Other explanations can be envisaged and deserve further exploration.  

 From the experimental point of view, the large statistical errors at the large negative $p_{\rm miss}$ setting for both shells will be reduced by combining our data with those of a recent measurement from MAMI on $^{12}$C at the same kinematic setting.  
The combined dataset could either improve the statistical significance of the deviations or show them to be statistical fluctuations.  
 
Furthermore, the study of $s$-shell knockout in future experiments may be useful in search of medium modifications of the nucleon form factors in nuclei.  Such a measurement has been proposed at MAMI on $^{40}$Ca \cite{40CaProposal}.  Since the $1s$-shell has large virtuality in $^{40}$Ca 
 (similar to that of $^{12}$C), it would be a suitable nucleus to extend the searches for medium modifications.

\begin{acknowledgments}
We would like to thank the Mainz Microtron operators and technical crew for the excellent operation of the accelerator. This work is supported by the Israel Science Foundation (Grants 390/15, 951/19) of the Israel Academy of Arts and Sciences, by the PAZY Foundation (Grant 294/18), by the Israel Ministry of Science, Technology and Space, by the Deutsche Forschungsgemeinschaft (Collaborative Research Center 1044), by the U.S. National Science Foundation (PHY-1205782).
We acknowledge the financial support from the Slovenian Research Agency (research core funding No.~P1\textendash 0102).  
\end{acknowledgments}
\vspace{0cm}
\bibliography{carbon}

\end{document}